\documentclass[12pt]{article}

\usepackage{graphicx}
\usepackage{pgfplots}
\usepackage{booktabs}
\usepackage{amsmath}
\usepackage{algorithm}
\usepackage{algorithmic}
\usepackage{amssymb}
\usepackage{amsfonts}
\usepackage{latexsym}
\usepackage{color}
\usepackage{subfigure}
\usepackage{latexsym}
\usepackage{graphicx,epstopdf}
\usepackage{subfigure}
\usepackage{authblk}
\usepackage{caption}
\usepackage{dsfont}

\newcommand{\be}{\begin{equation}}
\newcommand{\en}{\end{equation}}
\newcommand{\bea}{\begin{eqnarray}}
\newcommand{\ena}{\end{eqnarray}}
\newcommand{\beano}{\begin{eqnarray*}}
\newcommand{\enano}{\end{eqnarray*}}
\newcommand{\bee}{\begin{enumerate}}
\newcommand{\ene}{\end{enumerate}}

\newcommand{\Hil}{{\cal H}}

\newcommand{\Lc}{{\cal L}}

\newcommand{\Vc}{{\cal V}}

\newcommand{\Sc}{{\cal S}}

\newcommand{\1}{1 \!\! 1}

\begin{document}

\begin{center}
	{\Large \textbf{A note on a classical dynamical system and its quantization}} \vspace{2cm%
	}\\[0pt]
	
	{\large F. Bagarello}
	\vspace{3mm}\\[0pt]
	Dipartimento di Ingegneria,\\[0pt]
	Universit\`{a} di Palermo, I - 90128 Palermo,\\
	and I.N.F.N., Sezione di Catania\\
	E-mail: fabio.bagarello@unipa.it\\
	
	\vspace{7mm}

\end{center}

\vspace*{2cm}

\begin{abstract}
	\noindent In a recent paper a slightly modified version of the Bateman system, originally proposed to describe a damped harmonic oscillator, was proposed. This system is really different from the Bateman's one, in the sense that this latter cannot be recovered for any choice of its parameters. In this paper we consider this system and we show that, at a quantum level, it is not necessarily dissipative. In particular we show that the Hamiltonian of the system, when quantized, produces different behaviors, depending on some relations between its parameters. In fact, it gives rise to either a two dimensional (standard) harmonic oscillator, or to two independent oscillators, one of which is again {\em standard}, and a second one which is an inverted oscillator. The two cases are analyzed in terms of bosonic or pseudo-bosonic ladder operators, and the appearance of distributions for the inverted oscillator is commented.
\end{abstract}

\vspace*{1cm}


\vfill


\section{Introduction}\label{sec:intro}

The problem of quantizing dissipative systems, and the damped harmonic oscillator (DHO) in particular, has a long story. It is possible, for instance, to use Lagrangians which are explicitly time-dependent to recover the equation of motion for the DHO. This is the approach used, to cite just some authors, in  \cite{baldiotti,cheng,dekk,review2}. A different approach, originally proposed by Bateman and further studied by many authors along the years,  \cite{dekk,review2,bate,vitiello,dekk2,fesh}, is based on a time-independent Lagrangian which describes the DHO coupled with a fictitious amplified oscillator, which gains all the energy which is lost by the damped oscillator.

In recent years, inspired by \cite{nakano}, we have considered the problem of quantizing the Bateman Lagrangian in a series of papers, \cite{bag1,bag2,bgr,bag2023}, and among other results we have shown that it is possible to quantize the system using ladder operators of a special kind, the so-called {\em weak pseudo-bosons}, \cite{bagspringer}, whose natural framework is not that of Hilbert spaces, but rather that of distributions.

In this paper we consider a different system, recently proposed by Lee, \cite{lee}, in which the fictitious oscillator is replaced by a  (somehow) concrete oscillator coupled to another (real) oscillator. As we will show in a moment, the two oscillators are not {\em damped} and {\em amplified} in the same sense as the ones in \cite{bate}, and are coupled in a non trivial way. We should also mention that the system in \cite{lee} is different from the one considered recently in \cite{bag2023}, which on the other hand returns the Bateman system for a suitable choice of its parameters. In particular we will show that the classical system behaves in different ways depending on how the parameters of the system are fixed, and that this difference is observed also at the quantum level: the system may describe both conservative and dissipative systems. Indeed, at a quantum level, the same Hamiltonian gives rise either to  a two dimensional harmonic oscillator, or to two independent quantum oscillators, a {\em standard} oscillator in one direction, and an inverted oscillator in the {\em orthogonal} component. The eigenvalues of the Hamiltonian are real in the first case, but they turn out to be complex in the second situation.

The paper is organized as follows: in the next section we briefly introduce the classical system proposed in \cite{lee}, and we discuss some of its properties. In Section \ref{sect3} we quantize the system, and we consider two choices of its parameters in Section \ref{sect3a} and \ref{sect3b}. The first choice produces a two dimensional (standard) quantum harmonic oscillator, while the second choice produces a one dimensional harmonic oscillator coupled (but not interacting, as we will see!) with an inverted oscillator. We will show that several technical differences occur when dealing with these different choices. In particular in Section \ref{sect3a} we will need to work with a specific kind of non commutative quantum mechanics. while in Section \ref{sect3b} we will be forced to replace Hilbert spaces with tempered distributions. Our conclusions are given in Section \ref{sect4}.

\section{The classical system}\label{sect2}

The original Bateman system looks like the following system of differential equations:
\bea
\left\{
\begin{array}{ll}
	m\ddot x+\gamma \dot x+kx=0,\\
	m\ddot y-\gamma \dot y+ky=0,\\
\end{array}
\right.\label{21} \ena 
where, as stated, the variable $x(t)$ describes the time evolution of the DHO while $y(t)$ refers to the fictitious, amplified, harmonic oscillator (AHO).  The parameters $m,k$ and $\gamma$ are all positive. It is well known that these equations can be derived by the time-independent Lagrangian
\be L=m\dot x\dot y+\frac{\gamma}{2}(x\dot y-\dot xy)-kxy,
\label{22}\en
using the Euler-Lagrange equations $\frac{d}{dt}\frac{\partial L}{\partial \dot x}-\frac{\partial L}{\partial x}=0$ and $\frac{d}{dt}\frac{\partial L}{\partial \dot y}-\frac{\partial L}{\partial y}=0$. These equations, and their quantized version, have been considered by many authors in different ways. In particular, a no-go result has been deduced in \cite{bgr} stating that the ladder operators used by some authors to quantize the system, \cite{nakano}, do not admit any square-integrable vacuum. In fact, it can be proved that these vacua (there are two of them, \cite{bgr}) are indeed delta distributions.

The system (\ref{21}) was generalized in \cite{bag2023}, where the following differential equations were considered:
\be\label{23}
\left\{
\begin{array}{ll}
	m\ddot x+\gamma \dot x+kx=-2B\left(m\ddot y+ky\right)\\
	m\ddot y-\gamma \dot y+ky=-2A\left(m\ddot x+kx\right),\\
\end{array}
\right.
\en
which of course return those in (\ref{21}) when $A=B=0$. If $A$ or $B$ (or both) are non zero, then (\ref{23}) describes coupled oscillators, for which a Lagrangian can still be easily deduced and its quantization can be carried on, with similar problems as those in \cite{bgr}.

Recently, in \cite{lee}, a different system of differential equations has been considered in the context of Hamiltonian systems. The system is the following
\bea
\left\{
\begin{array}{ll}
	m\ddot x+\gamma \dot y+kx=-\lambda y,\\
	m\ddot y-\gamma \dot x+ky=-\lambda x.\\
\end{array}
\right.\label{24} \ena 
This describes two coupled oscillators in which the first, the one described by $x(t)$, is subjected to a force which depends on what the second oscillator does. Indeed, the force experienced by the $x$-oscillator is $F_x(t)=-\gamma \dot y(t)-\lambda y(t)$. Similarly, the second oscillator is driven by a force $F_y(t)=\gamma \dot x(t)-\lambda x(t)$. It is clear that the system in (\ref{24}) does not produce (\ref{21}) or (\ref{23}) for any choice of $\gamma$, $\lambda$ and $k$, all assumed to be real quantities. Hence this is really a different system, when compared with those considered in \cite{nakano,bgr,bag2023}, and for this reason it deserves our attention. From now on we will take $k=1$ to simplify the treatment, and we will only play with $\gamma$ and $\lambda$. It is easy to see that, using the Euler-Lagrange equations, the following Lagrangian
\be
\Lc=\frac{1}{2}\left(\dot x^2+\dot y^2\right)+(1+\gamma)\dot x\,y+x\dot y-\frac{1}{2}\left( x^2+ y^2\right)-\lambda \,x\,y
\label{25}\en
returns the equations in (\ref{24}) with $k=1$. From $\Lc$ we deduce the corresponding Hamiltonian as follows: we first compute $p_x=\frac{\partial \Lc}{\partial \dot x}=\dot x+(1+\gamma)y$ and $p_y=\frac{\partial \Lc}{\partial \dot y}=\dot y+x$, and then we put $\Hil=p_x\dot x+p_y\dot y-\Lc$:
\be
\Hil=\frac{1}{2}(p_x^2+2x^2)+\frac{1}{2}(p_y^2+(2+\gamma)y^2)-(1+\gamma)yp_x-xp_y+\lambda xy=\frac{1}{2}v^Thv,
\label{26}\en
where we have introduced
\be v=\begin{pmatrix}
	x \\
	y \\
	p_x\\
	p_y
\end{pmatrix}
\qquad \mbox{ and }\qquad 
h=\begin{pmatrix}
	2 & \lambda & 0 & -1\\
	\lambda & 2+\gamma & -(1+\gamma) & 0 \\
	0 & -(1+\gamma) & 1 & 0\\
	-1 & 0 & 0 & 1
\end{pmatrix}.
\label{27}\en
In (\ref{26}) $v^T$ is the transpose of $v$. We see that \be
tr(h)=6+\gamma, \qquad \det(h)=1-\gamma-\gamma^2-\lambda^2.\label{28}\en
It is also a simple exercise in linear algebra to diagonalize $h$, which is a symmetric matrix\footnote{Of course, it is also self-adjoint, due to the fact that $\lambda$ and $\gamma$ are real.}: let $S$ be a unitary matrix, $S^\dagger=S^{-1}$, such that
\be
h_d=ShS^{-1}=\begin{pmatrix}
	E_1 & 0 & 0 & 0\\
	0 & E_2 & 0 & 0 \\
	0 & 0 & E_3 & 0\\
	0 & 0 & 0 & E_4
\end{pmatrix},
\label{29}\en
where $E_j$ are the real eigenvalues of $h$. Of course we must have
$$
tr(h_d)=\sum_{j=1}^4E_j=tr(h)=6+\gamma,
$$
and
$$ \det(h_d)=\prod_{j=1}^4E_j=\det(h)=1-\gamma-\gamma^2-\lambda^2.
$$
If we further introduce
\be
\Vc=Sv=\begin{pmatrix}
	\tilde X \\
	\tilde Y \\
	\tilde P_X\\
	\tilde P_Y
\end{pmatrix},
\label{210}\en
then 
\be
\Hil=\frac{1}{2}\left(E_1\tilde X^2+E_2\tilde Y^2+E_3\tilde P_X^2+E_4\tilde P_Y^2\right),
\label{211}\en
which is diagonal and apparently looks like two uncoupled harmonic oscillators. Indeed, at least when quantizing the system, it might easily happen (and indeed this will be a possibility, as we will show in Section \ref{sect3a}) that $(\tilde X,\tilde P_X)$ do not {\em naturally} commute with $(\tilde Y,\tilde P_Y)$. Also, it may happen that not all the eigenvalues of $h$ are positive. In this particular case $\Hil$ describes an inverted harmonic oscillator (IHO), at least in one variable. This is exactly what we will see in Section \ref{sect3b}.

To better understand this aspect, we introduce the function $f(\gamma,\lambda)=\det(h)$ and we study its sign. Introducing the values $\gamma_\pm=\frac{1}{2}(-1\pm\sqrt{5})$, it is possible to check that $f(\gamma,\lambda)<0$ either if (Case 1) $\gamma<\gamma_-$ or $\gamma>\gamma_+$ for arbitrary (real) $\lambda$, or (Case 2) if $\gamma_-<\gamma<\gamma_+$, for $\lambda<-\sqrt{1-\gamma-\gamma^2}$ or $\lambda>\sqrt{1-\gamma-\gamma^2}$. In both these cases, recalling that $f(\gamma,\lambda)=\prod_{j=1}^4E_j$, it follows that one or three eigenvalues of $\Hil$ must necessarily be negative. This implies that, because of (\ref{211}), $\Hil$ cannot describe two standard harmonic oscillators, since in this case all the coefficients $E_j$ should be positive. 

\vspace{2mm}

{\bf Remark:--} It is interesting to observe that 
$$
p_\lambda(x,y,\dot x,\dot y)=\frac{1}{2}\left(\dot x^2+\dot y^2\right)+\frac{1}{2}\left( x^2+ y^2\right)+\lambda \,x\,y
$$
is an integral of motion for (\ref{24}), independently of the particular value of $\gamma$. Indeed we have, using (\ref{24}) and with simple computations, that $\dot p_\lambda(x,y,\dot x,\dot y)=0$. This means that the terms proportional to $\gamma$ in (\ref{24}) does not affect the nature of $p_\lambda(x,y,\dot x,\dot y)$, which is constant both in presence and in absence of these terms.

\section{Quantization}\label{sect3}

Quantizing a classical system is usually  not a trivial task, and for many reasons. First of all, there exist many different approaches to quantization. We refer to the very detailed review \cite{alieng}, where several possibilities are described, such as the Segal, the geometric, the canonical and the coherent state quantization, to list just some. In our paper we will only use the canonical quantization. But already at this level one should fix which are the classical variables to consider as {\em canonically conjugate}. In our particular case, the choice is between the pairs $(x,p_x)$ and $(y,p_y)$, or the new pairs $(\tilde X,\tilde P_X)$ and $(\tilde Y,\tilde P_Y)$. The latter choice would be more convenient, since $\Hil$ is already diagonal in these variables. However, we prefer the other choice which, in our opinion, is  more natural, since our original system is given in terms of $x$ and $y$. The use of $\tilde X$ and $\tilde Y$ is only a technicality to make computations easier, but the {\em relevant} variables, at least at this quantization level, are $x$, $y$, and their conjugate momenta $p_x$ and $p_y$. For this reason we now assume that the following canonical commutation relations hold (in units $\hbar=1$):
\be
[x,p_x]=[y,p_y]=i\1,
\label{31}\en
all the other commutators being zero. Here $\1$ is the identity operator on $\Lc^2(\mathbb{R}^2)$.
In what follows we will fix two different choices of $\lambda$ and $\gamma$ to cover the two different possibilities discussed before, i.e. the case in which all the eigenvalues of $\Hil$ are positive and the case in which they are not. Of course, other choices could be interesting but, in our opinion, they would not add much to what we will learn with the ones considered below.

\subsection{First choice: all positive eigenvalues}\label{sect3a}

We fix the values of our parameters as follows: $\lambda=\frac{1}{3}$ and $\gamma=-1$. Hence the system in (\ref{24}) looks like
$$
\left\{
\begin{array}{ll}
	\ddot x- \dot y+x=-\frac{y}{3},\\
	\ddot y+ \dot x+y=-\frac{x}{3},\\
\end{array}
\right.$$
recalling also that we always work with $k=1$. The matrix $h$ in (\ref{27}) is
$$
h=\begin{pmatrix}
	2 & \frac{1}{3} & 0 & -1\\
	\frac{1}{3} & 1 & 0 & 0 \\
	0 & 0 & 1 & 0\\
	-1 & 0 & 0 & 1
\end{pmatrix},
$$
and the eigenvalues are $E_1=\frac{8}{3}$, $E_2=E_3=1$ and $E_4=\frac{1}{3}$, with $\det(h)=\det(h_d)=\frac{8}{9}$. In this case the matrix $S$ in (\ref{29}) turns out to be
\be
S=\begin{pmatrix}
	-\sqrt{\frac{5}{7}} & -\sqrt{\frac{1}{35}} & 0 & \frac{3}{\sqrt{35}}\\
	0 &  \frac{3}{\sqrt{10}} & 0 & \frac{1}{\sqrt{10}} \\
	0 & 0 & 1 & 0\\
	\sqrt{\frac{2}{7}} & -\frac{1}{\sqrt{14}} & 0 &  \frac{3}{\sqrt{14}}
\end{pmatrix},
\label{32}\en
which is unitary, $S^\dagger=S^{-1}$, and from (\ref{210}) we deduce that
\be
\Vc=Sv=\begin{pmatrix}
	\tilde X \\
	\tilde Y \\
	\tilde P_X\\
	\tilde P_Y
\end{pmatrix}=\begin{pmatrix}
-\sqrt{\frac{5}{7}}\,x-\sqrt{\frac{1}{35}}\,y+\frac{3}{\sqrt{35}}\,p_y \\
\frac{3}{\sqrt{10}}\,y+\frac{1}{\sqrt{10}}\,p_y\\
p_x\\
\sqrt{\frac{2}{7}}\,x-\sqrt{\frac{1}{14}}\,y+\frac{3}{\sqrt{14}}\,p_y
\end{pmatrix}.
\label{33}\en
The Hamiltonian in (\ref{26}) is 
\be
\Hil=\frac{1}{2}(p_x^2+2x^2)+\frac{1}{2}(p_y^2+y^2)-xp_y+\frac{1}{3} xy=\frac{1}{2}\left(\tilde P_X^2+\frac{8}{3}\tilde X^2\right)+\frac{1}{2}\left(\frac{1}{3}\tilde P_Y^2+\tilde Y^2\right).
\label{34}\en
Using now the formulas in (\ref{31}) we deduce the following commutators for the {\em new} variables $\tilde X,\tilde Y,\tilde P_X$ and $\tilde P_Y$:
\be
[\tilde X,\tilde Y]=-i\sqrt{\frac{2}{7}}\,\1, \qquad [\tilde X,\tilde P_X]=-i\sqrt{\frac{5}{7}}\,\1, \qquad [\tilde X,\tilde P_Y]=0,
\label{35}\en
together with
\be
[\tilde Y,\tilde P_X]=0, \qquad [\tilde Y,\tilde P_Y]=i\sqrt{\frac{5}{7}}\,\1, \qquad [\tilde P_X\tilde ,P_Y]=-i\sqrt{\frac{2}{7}}\,\1.
\label{36}\en
In particular we see that both $[\tilde X,\tilde Y]$ and $[\tilde P_X,\tilde P_Y]$ are different from zero. Hence the new variables $\tilde X,\tilde Y,\tilde P_X$ and $\tilde P_Y$ obeys the commutation relations typical of non commutative quantum mechanics, \cite{ncqm1,ncqm2,ncqm3}, and the Hamiltonian in (\ref{25}) describes indeed a two-dimensional harmonic oscillator, but in a non commutative space. In particular, if we put 
$$
\Hil_{\tilde X}=\frac{1}{2}\left(\tilde P_X^2+\frac{8}{3}\tilde X^2\right), \qquad \Hil_{\tilde  Y}=\frac{1}{2}\left(\frac{1}{3}\tilde P_Y^2+\tilde Y^2\right),
$$ 
we see that $[\Hil_{\tilde X},\Hil_{\tilde Y}]\neq0$. The next step consists then in changing again variable adopting a Bopp shift to produce a different, but more {\em manageable}, Hamiltonian, \cite{bopp}.

The idea of this shift is to look for other variables, $(X,Y,P_X,P_Y)$, which are suitable linear combinations of the $(\tilde X,\tilde Y,\tilde P_X,\tilde P_Y)$  satisfying, first of all, the canonical commutation relations
\be
[X,P_X]=[Y,P_Y]=i\1,
\label{37}\en
all the other commutators being zero.

\vspace{2mm}

{\bf Remark:--} It should be emphasized that, as we will see explicitly, the $(X,Y,P_X,P_Y)$ are different from the original variables $(x,y,p_x,p_y)$. In other words, we are not simply going back to the original variables but we are going from two canonical pairs $(x,y,p_x,p_y)$ to two different canonical pairs   $(X,Y,P_X,P_Y)$ passing through two non-canonical pairs $(\tilde X,\tilde Y,\tilde P_X,\tilde P_Y)$. This will be useful since it will be possible to rewrite the Hamiltonian $\Hil$ in (\ref{34}) as the sum of two commuting harmonic oscillators, for which eigenvectors and eigenvalues can be easily constructed, see (\ref{310}) below.

\vspace{2mm}

We assume that the relation between  $(X,Y,P_X,P_Y)$ and $(\tilde X,\tilde Y,\tilde P_X,\tilde P_Y)$ is as follows:
\be
\tilde X=a_1X+b_1P_Y, \quad \tilde Y=a_2Y+b_2P_X, \quad \tilde P_X=a_3P_X+b_3Y, \quad \tilde P_Y=a_4P_Y+b_4X.   
\label{38}\en
Then, assuming (\ref{37}), formulas (\ref{35}) and (\ref{36}) are recovered if the above coefficients are chosen as follows:
\be
a_1=a_4=0, \quad a_2=\sqrt{\frac{2}{5}}\,b_3,\quad a_3=\sqrt{\frac{2}{7}}\,\frac{1}{b_4}, \quad b_1=\sqrt{\frac{5}{7}}\,\frac{1}{b_3}, \quad b_2=-\sqrt{\frac{5}{7}}\,\frac{1}{b_4}.
\label{39}\en
We see that these are $\infty^2$ solutions: different choices of non zero $b_3$ and $b_4$ produce different solutions and, therefore, different relations between $(X,Y,P_X,P_Y)$ and $(\tilde X,\tilde Y,\tilde P_X,\tilde P_Y)$. What is also extremely useful, as anticipated in the previous Remark, and was actually used to find the values in (\ref{39}), is that if we replace (\ref{38}) in $\Hil$, its expression in terms of the new variables is what we were looking for:
\be
\Hil=\Hil_X+\Hil_Y,
\label{310}\en
where
$$
\Hil_X=\frac{1}{2}\left(\frac{P_X^2}{b_4^2}+\frac{b_4^2\,X^2}{3}\right), \qquad \Hil_Y=\frac{1}{2}\left(\frac{40\,P_Y^2}{21\,b_3^2}+\frac{7\,b_3^2\,Y^2}{5}\right),
$$
which can be rewritten as 
\be
\Hil_X=\frac{1}{2}\left(P_X^2+\frac{X^2}{3}\right), \qquad \Hil_Y=\frac{1}{2}\left(P_Y^2+\frac{8Y^2}{3}\right),
\label{311}\en
if we further fix $b_4=1$ and $b_3=\sqrt{\frac{40}{21}}$. It is clear that, now, $[\Hil_X,\Hil_Y]=0$. So they can be diagonalized together, and we can find easily the eigenstates and the eigenvalues of $\Hil$. For that we introduce $\omega_X=\sqrt{\frac{1}{3}}$,  $\omega_Y=\sqrt{\frac{8}{3}}$, and the operators
\be
a_X=\frac{1}{\sqrt{2\omega_X}}\left(\omega_X\,X+iP_X\right), \qquad a_Y=\frac{1}{\sqrt{2\omega_Y}}\left(\omega_Y\,Y+iP_Y\right).
\label{312}\en
It follows that
\be
[a_X,a_X^\dagger]=[a_Y,a_Y^\dagger]=\1,
\label{313}\en
all the other commutator being zero. Moreover, $\Hil_X=\omega_X\left(a_X^\dagger a_X+\frac{1}{2}\1\right)$ and $\Hil_Y=\omega_Y\left(a_Y^\dagger a_Y+\frac{1}{2}\1\right)$. Next we introduce the vacua of $a_X$ and $a_Y$, $\xi_0(X)$ and $\eta_0(Y)$:
$
a_X\xi_0(X)=a_Y\eta_0(Y)=0$, and we further define $\xi_n(X)=\frac{(a_X^\dagger)^n}{\sqrt{n!}}\,\xi_0(X)$ and $\eta_m(Y)=\frac{(a_Y^\dagger)^m}{\sqrt{m!}}\,\eta_0(Y)$, $n,m\geq0$, and their tensor product 
\be
\varphi_{n,m}(X,Y)=\xi_n(X)\otimes\eta_m(Y)=\frac{1}{\sqrt{n!\,m!}}(a_X^\dagger)^n(a_Y^\dagger)^m\varphi_{0,0}(X,Y),
\label{314}\en
where $\varphi_{0,0}(X,Y)=\xi_0(X)\otimes\eta_0(Y)$ is the common vacuum of the operators $a_X\otimes\1_Y$ and $\1_X\otimes a_Y$, with obvious notation. It follows that
\be
\Hil\varphi_{n,m}=E_{n,m}\varphi_{n,m},
\label{315}\en
where $E_{n,m}=\omega_X\left(n+\frac{1}{2}\right)+\omega_Y\left(m+\frac{1}{2}\right)$. Incidentally we observe that all these eigenvalues are non degenerate, other than being, clearly, strictly positive. The conclusion of this analysis is that the choice  $\lambda=\frac{1}{3}$ and $\gamma=-1$ produces, after some changes of variables, a {\em standard} two-dimensional harmonic oscillator, living in an $\Lc^2(\mathbb{R}^2)$ space and in which the eigenfunctions in (\ref{314}) are the typical ones, written in terms of Hermite polynomials and gaussians as follows:
\be
\xi_n(X)=\frac{1}{\sqrt{2^n\,n!}}\left(\frac{\omega_X}{\pi}\right)^{1/4}H_n(\sqrt{\omega_X}\,X)e^{-\frac{1}{2}\omega_XX^2},
\label{316}\en
and similarly
\be
\eta_m(Y)=\frac{1}{\sqrt{2^m\,m!}}\left(\frac{\omega_Y}{\pi}\right)^{1/4}H_m(\sqrt{\omega_Y}\,Y)e^{-\frac{1}{2}\omega_YY^2}.
\label{317}\en

\subsection{Second choice: one negative eigenvalue}\label{sect3b}

Let us now consider the following values of $\lambda$ and $\gamma$: $\lambda=\gamma=1$. Then the system (\ref{24}) looks like
$$
\left\{
\begin{array}{ll}
	\ddot x+ \dot y+x=-y,\\
	\ddot y- \dot x+y=-x,\\
\end{array}
\right.$$
recalling once more that we have fixed $k=1$ in (\ref{24}). The matrix $h$ in (\ref{27}) is now
$$
h=\begin{pmatrix}
	2 & 1 & 0 & -1\\
	1 & 3 & -2 & 0 \\
	0 & -2 & 1 & 0\\
	-1 & 0 & 0 & 1
\end{pmatrix},
$$
and the eigenvalues are $E_1=\frac{1}{2}(5+\sqrt{17})$, $E_2=1+\sqrt{2}$, $E_3=\frac{1}{2}(5-\sqrt{17})$ and $E_4=1-\sqrt{2}$, with $\det(h)=\det(h_d)=-2$. We see that with this choice one eigenvalue of $h$ is negative, while all the others are positive. In this case the matrix $S$ in (\ref{29}) is slightly more complicated to write:
$$
S=\begin{pmatrix}
	-\sqrt{\frac{1}{102}(17-\sqrt{17})} & -\sqrt{\frac{1}{3}+\frac{4}{3\sqrt{17}}} & \sqrt{\frac{1}{102}(17+\sqrt{17})} & \frac{1}{\sqrt{51+12\sqrt{17}}}\\
	-\sqrt{\frac{1}{6}(2+\sqrt{2})} &  \frac{1}{2}\sqrt{\frac{1}{3}(2-\sqrt{2})} & -\sqrt{\frac{1}{6}(2-\sqrt{2})} & \frac{1}{\sqrt{12-6\sqrt{2}}} \\
	\sqrt{\frac{1}{102}(17+\sqrt{17})} & \sqrt{\frac{1}{3}-\frac{4}{3\sqrt{17}}} & \sqrt{\frac{1}{6}-\frac{1}{6\sqrt{17}}} & \frac{1}{\sqrt{51-12\sqrt{17}}}\\
	\frac{1}{\sqrt{3(2+\sqrt{2})}} & - \frac{1}{2}\sqrt{\frac{1}{3}(2+\sqrt{2})} & -\sqrt{\frac{1}{6}(2+\sqrt{2})} &  \frac{1}{\sqrt{12+6\sqrt{2}}}
\end{pmatrix},
$$
which is unitary, $S^\dagger=S^{-1}$, and from (\ref{210}) we deduce that
\be
\Vc=Sv=\begin{pmatrix}
	\tilde X \\
	\tilde Y \\
	\tilde P_X\\
	\tilde P_Y
\end{pmatrix}=\begin{pmatrix}
	a_1x+a_2y+a_3p_x+a_4p_y\\
	b_1x+b_2y+b_3p_x+b_4p_y\\
	c_1x+c_2y+c_3p_x+c_4p_y\\
	d_1x+d_2y+d_3p_x+d_4p_y\\
\end{pmatrix},
\label{318}\en
and where the coefficients are the following:
$$
a_1=\frac{-(3+\sqrt{17})}{2\sqrt{51+12\sqrt{17}}}, \quad a_2=\frac{-(4+\sqrt{17})}{\sqrt{51+12\sqrt{17}}}, \quad a_3=\frac{5+\sqrt{17}}{2\sqrt{51+12\sqrt{17}}}, 
$$
$$
a_4=\frac{1}{\sqrt{51+12\sqrt{17}}}, \quad b_1=- \frac{1}{2}\sqrt{\frac{2}{3}(2+\sqrt{2})}, \quad b_2= \frac{\sqrt{2}-1}{2}\sqrt{\frac{1}{3}(2+\sqrt{2})},
$$
$$
b_3= \frac{1-\sqrt{2}}{2}\sqrt{\frac{2}{3}(2+\sqrt{2})}, \quad b_4= \frac{1}{2}\sqrt{\frac{1}{3}(2+\sqrt{2})}, \quad c_1=\frac{-3+\sqrt{17}}{2\sqrt{51-12\sqrt{17}}},
$$
$$
c_2=\frac{-4+\sqrt{17}}{\sqrt{51-12\sqrt{17}}}, \quad c_3=\frac{5-\sqrt{17}}{2\sqrt{51-12\sqrt{17}}}, \quad c_4=\frac{1}{\sqrt{51-12\sqrt{17}}},
$$
$$
d_1=\frac{1}{\sqrt{3(2+\sqrt{2})}}, \quad d_2=\frac{-(\sqrt{2}+1)}{\sqrt{6(2+\sqrt{2})}}, \quad d_3=\frac{-(\sqrt{2}+1)}{\sqrt{3(2+\sqrt{2})}},
$$
$$
d_4=\frac{1}{\sqrt{6(2+\sqrt{2})}}.
$$
The Hamiltonian in (\ref{26}), for our choice of parameters, is now 
$$
\Hil=\frac{1}{2}(p_x^2+2x^2)+\frac{1}{2}(p_y^2+3y^2)-xp_y-2yp_x+ xy=$$
\be =\frac{1}{2}\left(\frac{5-\sqrt{17}}{2}\tilde P_X^2+\frac{5+\sqrt{17}}{2}\tilde X^2\right)+\frac{1}{2}\left((1-\sqrt{2})\tilde P_Y^2+(1+\sqrt{2})\tilde Y^2\right).
\label{319}\en
Assuming now the commutators in (\ref{31}) it is quite surprising to see that, despite the very complicated expression of the transformation, we find that
\be
[\tilde X,\tilde Y]=[\tilde X,\tilde P_Y]=[\tilde X,\tilde P_Y]=[\tilde P_X,\tilde P_Y]=0,
\label{320}\en
and
\be
[\tilde X,\tilde P_X]=-i\1, \qquad [\tilde Y,\tilde P_Y]=i\1.
\label{321}\en
On other words, despite its apparent difficulty, this case is simpler than the previous one, at least for what concerns  the diagonalization of the Hamiltonian $\Hil$. Renaming the above operators as in
\be
X=\tilde P_X, \quad P_X=\tilde X, \quad Y=\tilde Y, \quad P_Y=\tilde P_Y,
\label{322}\en
it follows that these are canonically conjugate, in the sense that they obey the same commutation relations as in (\ref{37}): $[X,P_X]=[Y,P_Y]=i\1$,
all the other commutators being zero, without the need of considering any further transformation (like the Bopp shift considered before). With these definitions we can rewrite
\be
\Hil=\Hil_X-\Hil_Y,
\label{323}\en
where
$$
\Hil_X=\frac{1}{4}\left((5+\sqrt{17}) P_X^2+(5-\sqrt{17}) X^2\right), 
$$
and
$$
\Hil_Y=\frac{1}{2}\left((\sqrt{2}-1) P_Y^2-(\sqrt{2}+1) Y^2\right).
$$
We stress the presence of the minus sign in (\ref{323}). It is clear that $[\Hil_X,\Hil_Y]=0$, so that the eigenvectors of $\Hil$ can be deduced diagonalizing simultaneously $\Hil_X$ and $\Hil_Y$. 

The analysis of $\Hil_X$ is not particularly different from what we have seen before, for the previous case. If we put $\omega_X=\sqrt{8}$, $\alpha=\sqrt{\frac{2\sqrt{2}}{5+\sqrt{17}}}$ and
$$
a_X=\frac{1}{\sqrt{2}}\left(\alpha X+i\frac{1}{\alpha}P_X\right),
$$
then $[a_X,a_X^\dagger]=\1$ and $\Hil_X=\omega_X\left(a_X^\dagger a_X+\frac{1}{2}\1\right)$. Hence we can proceed as in Section \ref{sect3a}:   we introduce the vacuum of $a_X$, which we again call $\xi_0(X)$, 
$
a_X\xi_0(X)=0$, and we further define $\xi_n(X)=\frac{(a_X^\dagger)^n}{\sqrt{n!}}\,\xi_0(X)$. These functions do not differ significantly from those in (\ref{316}), except that for some minor details which are not really interesting for us here.

The situation is different when we consider $\Hil_Y$. We can rewrite 
\be
\Hil_Y=\frac{\sqrt{2}-1}{2}\left(P_Y^2-\frac{\sqrt{2}+1}{\sqrt{2}-1} Y^2\right)=(\sqrt{2}-1)\Hil_{IO},
\label{324}\en
where $\Hil_{IO}=\frac{1}{2}\left(P_Y^2-\Omega^2 Y^2\right)$, with $\Omega^2=\frac{\sqrt{2}+1}{\sqrt{2}-1}$. We see that $\Hil_{IO}$ is the Hamiltonian of an inverted harmonic oscillator, for which all the results deduced in \cite{bag2022} can be adopted. In particular, it is shown in \cite{bag2022} that the natural functional space to consider in connection with an inverted oscillator is not an $\Lc^2$-space, but it is rather the space of tempered distributions, $\Sc'(\mathbb{R})$, \cite{reed}. It was also discussed that weak pseudo-bosons, \cite{bagspringer}, are relevant in its analysis, while ordinary bosons are not. We avoid here all the mathematical details related to the distributional nature of the (generalized) eigenstates of $\Hil_{IO}$, for which we refer to \cite{bag2022}, where the inverted oscillator is analyzed as a {\em weak limit} of a Swanson-like Hamiltonian, \cite{swan}. We only focus on the construction of its (again, generalized) eigenstates and on the related eigenvalues. For that we introduce the operators
\be
A=\frac{1}{\sqrt{2\Omega}}\left(e^{i\pi/4}\Omega Y+ie^{-i\pi/4}P_Y\right), \qquad B=\frac{1}{\sqrt{2\Omega}}\left(e^{i\pi/4}\Omega Y-ie^{-i\pi/4}P_Y\right).
\label{325}\en
$A$ and $B$ are densely defined on $\Sc(\mathbb{R})$, the space of the test functions, which is dense in $\Lc^2(\mathbb{R})$ and which is stable under the action of $A$, $B$ and of their adjoints, \cite{bagspringer}. It is clear that $B$ is not the adjoint of $A$, $B\neq A^\dagger$. We also have that
\be
[A,B]\Phi(Y)=\Phi(Y),
\label{326}\en
for all $\Phi(Y)\in\Sc(\mathbb{R})$. In \cite{bag2022} it is discussed how these operators, together with the commutator in (\ref{326}), can be extended to the set $\Sc'(\mathbb{R})$. Using the weak pseudo-bosonic approach we can look for a distribution $\eta_0(Y)$ such that $A\eta_0(Y)=0$. Then we define $\eta_n(Y)=\frac{B^n}{\sqrt{n!}}\,\eta_0(Y)$. We know, \cite{bag2022}, that these are not square integrable functions but tempered distributions, whose explicitly expression can be found in \cite{bag2022}, and it involves again Hermite polynomials and exponential (which look like gaussian, but with a purely imaginary argument!) functions. We also know that $N\eta_n=n\eta_n$, $n\geq0$, where $N=BA$. This equality makes sense since $\Sc'(\mathbb{R})$ is stable under the action of both $A$ and $B$, and since $\eta_0(Y)\in\Sc'(\mathbb{R})$. Next, since $\Hil_{IO}=i\Omega\left(BA+\frac{1}{2}\1\right)$, we conclude that $\Hil_{IO}\eta_n(Y)=i\Omega\left(n+\frac{1}{2}\right)\eta_n(Y)$: the eigenvalues of $\Hil_{IO}$ are purely imaginary. The same conclusions holds for $\Hil_Y$ in (\ref{324}):
\be
\Hil_Y\eta_n(Y)=i\left(n+\frac{1}{2}\right)\eta_n(Y),
\label{327}\en
for all $n\geq0$.

\vspace{2mm}

{\bf Remark:--} Notice that our construction does not exclude the possibility that other parts of the spectrum of $\Hil_{IO}$ exist, which could belong to its continuous or to its discrete spectrum. Our strategy might give only part of the spectrum. Moreover, it should be emphasized that what we are considering here is not really the spectrum in an ordinary sense, which strongly refers  to the fact of working in Hilbert spaces. That's why we have often used before the word "generalized".

\vspace{2mm}

Going back to $\Hil=\Hil_X-\Hil_Y$ it is clear that its (generalized) eigenstates are $\varphi_{n,m}(X,Y)=\xi_n(X)\otimes\eta_m(Y)$, $n,m\geq0$, and that
\be
\Hil\varphi_{n,m}(X,Y)=E_{n,m}\varphi_{n,m}(X,Y),
\label{328}\en
for all $n,m\geq0$, where 
$$
E_{n,m}=\omega_X\left(n+\frac{1}{2}\right)+i\left(m+\frac{1}{2}\right),
$$
which has always a non zero imaginary part. This is not in contrast with the fact that $\Hil$ is formally self-adjoint, see (\ref{319}): formal and rigorous self-adjointness are quite different, in fact, \cite{reed2}. What this model suggests, we believe, is that the unboundedness of the operators intrinsically connected with the system does not always permit a {\em canonical treatment} in $\Lc^2(\mathbb{R}^2)$. Sometimes, as in this case, we have to enlarge the functional space which is needed for our analysis.

\section{Conclusions}\label{sect4}

We have considered a classical dynamical system recently proposed in the literature as the basis to discuss its quantization and we have shown that different choices of its parameters give rise to different behavior of the quantized system. In particular we have shown that the same classical system can produce a two dimensional quantum oscillator living in a $\Lc^2(\mathbb{R}^2)$ space, and a different system consisting of an ordinary harmonic oscillator in one variable and an IHO in another variable, for which the role of square-integrable functions is not essential, and for which a distributional approach appears to be more useful.

The analysis carried out in this paper suggests, together with similar results deduced in previous papers, that the role of distributions in specific quantum systems, and in particular in non conservative systems, may be relevant for their rigorous analysis.

\section*{Acknowledgements}
The author acknowledges partial financial support from Palermo University and from G.N.F.M. of the INdAM. This work has also been partially supported by the PRIN grant {\em Transport phenomena in low dimensional
	structures: models, simulations and theoretical aspects}- project code 2022TMW2PY - CUP B53D23009500006, and partially by  the PRIN-PNRR 2022, “CAESAR” - FAIR PE0000013, CUP J53C22003010006 and by project ICON-Q, Partenariato Esteso NQSTI - PE00000023, Spoke 2.

\end{document}